\documentclass[english,twocolumn,prb,amssymb,amsmath,superscriptaddress]{revtex4}
\usepackage[latin9]{inputenc}
\usepackage{graphicx,color}
\usepackage[bookmarks]{hyperref}
\usepackage{babel}

\newcommand{\be}{\begin{equation}}
\newcommand{\ee}{\end{equation}}
\newcommand{\bea}{\begin{eqnarray}}
\newcommand{\eea}{\end{eqnarray}}
\newcommand{\ua}{\uparrow}
\newcommand{\da}{\downarrow}
\newcommand{\sbold}{\mathbf{s}}
\newcommand{\sboldbar}{\overline{\mathbf{s}}}

\newcommand{\ue}{\ensuremath{{\rm{e}}}}
\newcommand{\teo}{\ensuremath{{\mathcal{U}}}}
\newcommand{\initconf}{\ensuremath{{0}}}

\newcommand{\RG}{\ensuremath{{(\rm{r})}}}

\definecolor{mygreen}{rgb}{0,0.5,0}
\definecolor{myblue}{rgb}{0,0,0.75}
\definecolor{mymagenta}{cmyk}{0,1,0,0.12}

\usepackage{umoline}

\begin{document}

\title{Many-body localization and quantum ergodicity in disordered long-range Ising models}
\author{Philipp Hauke}
\affiliation{Institute for Quantum Optics and Quantum Information of the Austrian Academy of Sciences, 6020 Innsbruck, Austria}
\affiliation{Institute for Theoretical Physics, University of Innsbruck, 6020 Innsbruck, Austria}
\author{Markus Heyl}
\affiliation{Institute for Quantum Optics and Quantum Information of the Austrian Academy of Sciences, 6020 Innsbruck, Austria}
\affiliation{Institute for Theoretical Physics, University of Innsbruck, 6020 Innsbruck, Austria}

\begin{abstract}

Ergodicity in quantum many-body systems is---despite its fundamental importance---still an open problem. 
Many-body localization provides a general framework for quantum ergodicity, and may therefore offer important insights. 
However, the characterization of many-body localization through simple observables is a difficult task. 
In this article, we introduce a measure for distances in Hilbert space for spin-1/2 systems that can be interpreted as a generalization of the Anderson localization length to the many-body Hilbert space. We show that this many-body localization length is equivalent to a simple local observable in real space, which can be measured in experiments of superconducting qubits, polar molecules, Rydberg atoms, and trapped ions.  
Using the many-body localization length and a necessary criterion for ergodicity that it provides, we study many-body localization and quantum ergodicity in power-law-interacting Ising models subject to disorder in the transverse field. 
Based on the nonequilibrium dynamical renormalization group, numerically exact diagonalization, and an analysis of the statistics of resonances we find a many-body localized phase at infinite temperature for small power-law exponents. Within the applicability of these methods, we find no indications of a delocalization transition. 
 
\end{abstract}

\maketitle
\date{\today}


\section{Introduction}
\label{sev:Intro}

Ergodicity is a fundamental concept of statistical physics. If a classical system is ergodic, phase-space trajectories cover uniformly constant energy hyper-surfaces, such that time and microcanonical ensemble averages become equivalent \cite{Eckmann1985bn}. 
Although attempts to extend these ideas to the quantum regime date back to von Neumann's quantum ergodic theorem  \cite{Neumann1929,Goldstein2010mb}, a general conceptual understanding of quantum ergodicity has not yet been achieved \cite{Polkovnikov2011kx}. This, however, is crucial for fundamental questions such as regarding the thermalization of closed quantum many-body systems. 
A lack of quantum ergodicity can, in analogy to the classical phase-space description, be seen as localization in Hilbert space, for which a general framework has been introduced recently: many-body localization(MBL) \cite{Shepelyansky1994,Altshuler1997hx,Basko2006}. 

MBL phases exhibit further peculiar properties~\cite{Nandkishore2015,Altman2014} beyond the fundamental question of quantum ergodicity. Perhaps most notable among them is a universal temporal growth of entanglement following global quenches out of weakly entangled initial states \cite{Znidaric2008,Bardarson2012,Vosk2012,Serbyn2013,Vosk2013}. Additionally, MBL phases can exhibit finite-temperature phase transitions even in one dimension \cite{Basko2006,Aleiner2010}, which are excluded for thermodynamic phases. Even more, many-body localization can stabilize order in one dimension over the full spectrum \cite{Huse2013,Bahri2013,Chandran2014}, which may be of interest for designing quantum-information devices \cite{Nandkishore2015}. 
Compared to conventional localization in real space \cite{Anderson1958wx,Abrahams1979}, however, theoretical calculations of many-body localization suffer from the complexity of the underlying geometry---the many-body Hilbert space. Therefore, revealing many-body localization properties and finding suitable, experimentally accessible quantities for their characterization remains challenging. 

In this article, we introduce an observable that measures distances in Hilbert space, and as such can be interpreted as a many-body generalization of the Anderson localization length. 
Importantly, it can be obtained via simple local measurements such as on-site magnetizations. This observable thus opens a feasible and efficient route for studying many-body localization in experiments. 
Using this measure, we study in detail the disordered quantum Ising model with power-law interactions at small power-law exponent $\alpha \leq 1$, which is relevant to experiments on polar molecules, Rydberg atoms, superconducting qubits, and trapped ions. Our calculations predict that transverse-field disorder drives the model into a MBL phase even  at infinite temperature.
These findings are drawn from the recently introduced nonequilibrium dynamical renormalization group (ndRG) \cite{Heyl2013}, from extensive numerical simulations using exact diagonalization (ED), as well as  from an analysis of the statistics of resonant Hilbert-space configurations. 
Within the applicability of these methods, we find no indications of a delocalization transition, even for weak disorder strength. 

The remainder of this paper is organized as follows. In Sec.~\ref{sec:LRI} we introduce the disordered quantum Ising chain with power-law interactions, which we use to illustrate our considerations. Basic principles of many-body localization are discussed in Sec.~\ref{sec:MBL}, including the main result of this work, the many-body localization length. In Sec.~\ref{sec:results}, we calculate the many-body localization length for the disordered power-law-interacting Ising chain at infinite temperature, using the ndRG as well as extensive numerically exact simulations, indicating a many-body localized phase at nonzero disorder strength. We corroborate these predictions by an analytical analysis of the statistics of resonances.

\section{Long-range Ising chains}
\label{sec:LRI}

In this work, we study localization beyond the single-particle, i.e., Anderson-localized, limit,  by considering disordered Ising chains with algebraic long-range couplings between the spins, 
\be
	H_\mathrm{Ising} = \sum_{l<m} \frac{\mathcal{J}}{|l-m|^\alpha} \sigma_l^x \sigma_m^x + \sum_{l=1}^N h_l \sigma_l^z,
\label{eq:defHamiltonianIsing}
\ee
with $\sigma_l^\mu$, $\mu=x,y,z$, the Pauli matrices, and where the exponent $\alpha\geq 0$ determines the range of the interactions. 
This class of Ising models appear in many natural contexts \cite{Campa2008}---especially in systems with Coulomb, dipole--dipole, and van-der-Waals interactions---and they can be synthesized in a variety of architectures, including trapped ions \cite{Friedenauer2008,Britton2012,Richerme2014,Jurcevic2014}, superconducting qubits \cite{Linzen2007,Niskanen2007,DiCarlo2009,Boixo2014}, polar molecules \cite{Lahaye2009,Carr2009,Trefzger2011}, and Rydberg atoms \cite{Saffman2010,Loew2012}. 
To connect to current experiments, we choose  antiferromagnetic interactions ($\mathcal{J}>0$) and open boundary conditions, although our main results do not dependent on these choices. The transverse fields $h_l\in[-W,W]$ are drawn from uncorrelated uniform distributions.

In order to assure extensivity of the full many-body spectrum of the Hamiltonian~(\ref{eq:defHamiltonianIsing}), we follow the Kac prescription \cite{Kac1963} and normalize the coupling constant by 
\be
	\mathcal{J} = \frac{J}{\mathcal{N}(\alpha)}, \quad \mathcal{N}(\alpha) = \frac{1}{N-1} \sum_{l\not=m} \frac{1}{|l-m|^\alpha}.
\label{eq:couplingScaling}
\ee
This choice reproduces the $N^{-1}$ scaling of $\mathcal{J}$ in the fully connected model at $\alpha=0$ (see Ref.~\onlinecite{Ioffe2010}) as well as the system-size independence for $\alpha>1$.

In equilibrium, the transverse-field Ising model, Eq.~(\ref{eq:defHamiltonianIsing}), hosts paramagnetic und magnetically ordered phases, both for the clean \cite{Koffel2012} and the disordered system \cite{Ioffe2010,Juhasz2014}. Specifically, the Ising critical point is unstable against disorder for $\alpha>1$ and the magnetic quantum phase transition is governed by a strong-disorder fixed point~\cite{Juhasz2014} with magnetic order only at vanishing temperature. For the fully-connected case at $\alpha=0$ the magnetic phase extends also to non-zero temperatures, with a phase boundary as determined in Ref.~\onlinecite{Ioffe2010}.

Localization properties of the disordered long-range Ising models in Eq.~(\ref{eq:defHamiltonianIsing}) and related systems have already been studied in the literature. First of all, in the limit $\alpha \to \infty$ where the long-range model reduces to an exactly solvable nearest-neighbor Ising chain, the system becomes an Anderson insulator at nonzero disorder strength. At finite $\alpha<\infty$, however, the situation is not completely clear. In particular, it has been argued that in the regime $1<\alpha<2$, the system delocalizes at any finite disorder strength~\cite{Burin2006}. For $\alpha>2$ a many-body localization transition might be possible as observed for related long-range XXZ chains, it has, however, not been explicitly shown yet~\cite{Yao2013,Burin2014}. For the infinitely connected limit with $\alpha=0$ on the other hand, analytical calculations have revealed a MBL phase for nonvanishing disorder strength~\cite{Ioffe2010}. For the regime $0<\alpha<=1$, the situation is much less clear. It is one purpose of this work to show that the MBL phase at any nonzero disorder strength in the long-range Ising chains extends from $\alpha=0$ to the entire regime $\alpha<=1$.

\section{Many-body localization length}
\label{sec:MBL}

In this section, we discuss in more detail many-body localization and its connection to quantum ergodicity. In particular, we will explicitly show the mapping of interacting spin models, such as the Hamiltonian in Eq.~(\ref{eq:defHamiltonianIsing}), onto noninteracting Anderson models on a complex graph of spin configurations. We will then present the main result of this work, a definition of a  distance in this complex graph for spin-1/2 models, which can be interpreted as the many-body localization length and which is experimentally accessible.

\subsection{Many-body localization and quantum ergodicity}

Quantum ergodicity can be viewed from a dynamical or a static perspective. Dynamically, quantum ergodicity implies thermalization. The long-time values of local and quasi-local observables after a nonequilibrium evolution coincide with those of a thermal ensemble for almost any initial condition, because time and ensemble averages are equivalent. There is, however, one particular situation where ergodicity is not sufficient for thermalization, but rather requires an additional principle~\cite{Heyl2013}: Whenever the asymptotic long-time state of a system, when thermalized, lies in a symmetry-broken phase of the model, a dynamical symmetry breaking has to occur restricting the long-time dynamics to one symmetry-broken sector.
The Eigenstate-Thermalization-Hypothesis (ETH) has been conjectured as an underlying principle for thermalization in closed quantum many-body systems~\cite{Deutsch1991fe,Srednicki1994eg,Tasaki1998ww,Rigol2008vm}: If ETH holds for a given system, then it thermalizes. However, the connection between microscopic details of a system and the applicability of ETH is still not fully clarified.  
Note that in this article we do not distinguish between ergodicity and mixing \cite{Eckmann1985bn}, because the observables under study approach stationary values during time evolution, so long-time averages (ergodicity) and asymptotic long-time values (mixing) coincide. 

From a static point of view, quantum ergodicity can be associated with delocalization in Hilbert space \cite{Altshuler1997hx}. Let $|\mathbf{s}\rangle =|s_1,\dots,s_N\rangle$, with $|s_l\rangle=\left|\ua\right\rangle,\left| \da\right\rangle$, be an arbitrary spin configuration in the $\sigma^z$ basis, i.e., an eigenstate of the Hamiltonian Eq.~(\ref{eq:defHamiltonianIsing}) at $\mathcal{J}=0$. Adiabatically turning on the coupling $\mathcal{J}$ deforms the eigenstates and mixes different spin configurations. When each spin configuration only acquires weak perturbative corrections, the system will remain localized in Hilbert space around the $\mathcal{J}=0$ eigenstates and will therefore not be ergodic. 
Delocalization, on the other hand, is driven by the proliferation of resonances between configurations in Hilbert space. 

\subsection{Mapping onto Anderson model on a complex graph}
\label{ssec:AndersonModel}

With interactions beyond nearest neighbours, the Ising model in Eq.~(\ref{eq:defHamiltonianIsing}) is not of single-particle type. But still, following Refs.~\onlinecite{Altshuler1997hx,Basko2006}, a mapping to a noninteracting (albeit complex) Anderson model is possible if we represent the Hilbert space by a lattice where each site is associated with one spin configuration $|\mathbf{s}\rangle$. The Ising model then finds an exact mapping to 
\be
	H_\mathrm{Ising}  = \sum_{\sbold} E_\sbold |\sbold \rangle \langle \sbold |  + \sum_{ \sbold,\sboldbar} V_{\sbold,\sboldbar} |\sboldbar \rangle \langle \sbold |,
\label{eq:AndersonModel}
\ee
i.e., an Anderson model on a complex graph with on-site energies $E_\sbold = \sum_l h_l s_l$. The Ising interaction couples all states that differ by two spin flips, inducing a hopping with amplitude $V_{\sbold,\sboldbar} = \langle \sbold | V | \sboldbar \rangle$, where $V=\sum_{l\not=m} \mathcal{J}_{lm}\sigma_l^x \sigma_m^x$ and $\mathcal{J}_{lm} =\mathcal{J}/|l-m|^{\alpha}$. Although the Hamiltonian in Eq.~(\ref{eq:AndersonModel}) is now noninteracting, the problem is still hard to solve due to the complexity of the underlying graph.
In particular, the hopping in the lattice of spin configurations is characterized by an unconventionally high connectivity, i.e., the number of lattice sites accessible by a single hopping process from a given site. Compared to the real-space problem, the connectivity is enhanced by a factor proportional to $N$. For example, the variable-range Ising chain has a connectivity in Hilbert space of $N(N-1)/2$, in contrast to $N-1$ in real space.

In the configurational space, one can define a distance $d(\sbold,\sboldbar)$ between two sites $|\sbold\rangle$ and $|\sboldbar\rangle$ by counting the number of spins that differ between the two configurations \cite{Altshuler1997hx} (Hamming distance). Fixing one site $|\sbold_\initconf\rangle$, the remaining lattice can be classified by grouping configurations of equal distance to $|\sbold_\initconf\rangle$ into `generations'. We define generation 1 as those states with  $d(\sbold,\sbold_\initconf)=2$, generation 2 those with $d(\sbold,\sbold_\initconf)=4$, up to $d(\sbold,\sbold_\initconf)=N$.

\subsection{Many-body localization length}
\label{ssec:MBLlength}

A good way to characterize localization of an Anderson insulator is by monitoring the spread of an initially localized wave function over time. In our case, an analog approach amounts to initializing the system in a `root' configuration $|\sbold_\initconf\rangle$, the most localized object in our graph, and studying how the mean distance from this initial site,
\be
      \mathcal{D}_{\sbold_\initconf}(t) = \sum_\sbold d(\sbold,\sbold_\initconf) P(\sbold,t),
\ee
increases during time evolution. Here, $P(\sbold,t)$ is determined by $P(\sbold,t)=|\langle \sbold|\sbold_\initconf(t)\rangle|^2$, the probability for the system to be in the configuration $|\sbold\rangle$, where  $|\sbold_\initconf(t)\rangle = \teo(t)|\sbold_\initconf\rangle$ is the initial configuration after time evolution under $\teo(t) = \exp(-i H_\mathrm{Ising}t)$.

The challenge is to measure $\mathcal{D}_{\sbold_\initconf}(t)$ in practice. As a major result of this work, this global quantity, characterizing the wave function in Hilbert space, is related to a local real-space autocorrelation function $\chi_{\sbold_\initconf}(t)$ via 
\be
	\mathcal{D}_{\sbold_\initconf}(t) = \frac{N}{2}\left[ 1 - \chi_{\sbold_\initconf}(t) \right]\,, 
\label{eq:defLocalizationLength}
\ee
with
\be
	\chi_{\sbold_\initconf}(t) =\frac{1}{N} \sum_{l=1}^N \left\langle \sbold_\initconf| \sigma_l^z(t) \sigma_l^z |\sbold_\initconf \right\rangle\,,
\label{eq:defLocalizationLength_chi}
\ee
where $\sigma_l^z(t) = \teo^\dag(t) \sigma_l^z \teo(t)$. One arrives at this result from the definition of $\mathcal{D}_{\sbold_\initconf}(t)$ when using $d(\sbold,\sbold_\initconf)=\sum_l (s_l - s_l^\initconf)^2/4$, with $s_l=+ 1,-1$ for $\left|{s_l}\right\rangle=\left|\ua\right\rangle,\left|\da\right\rangle$.  
The Hilbert space property $\mathcal{D}_{\sbold_\initconf}(t)$ can therefore be obtained from purely local measurements in \emph{real} space, provided the initial configuration is known. In the context of the Richardson model, a similar relation has been obtained recently, which, however, is restricted to particular initial states and the asymptotic long-time regime \cite{Buccheri2011}. Our Eq.~\eqref{eq:defLocalizationLength} is completely general and independent of the specific spin system. The local memory $\chi(t)$ is well known in the context of Anderson~\cite{Anderson1958wx} and many-body localization~\cite{Iyer2012}. Equation~\eqref{eq:defLocalizationLength} shows that in the many-body context it has a further important meaning by being related to distances in Hilbert space.

\begin{figure}
\centering
\includegraphics[width = 1\columnwidth]{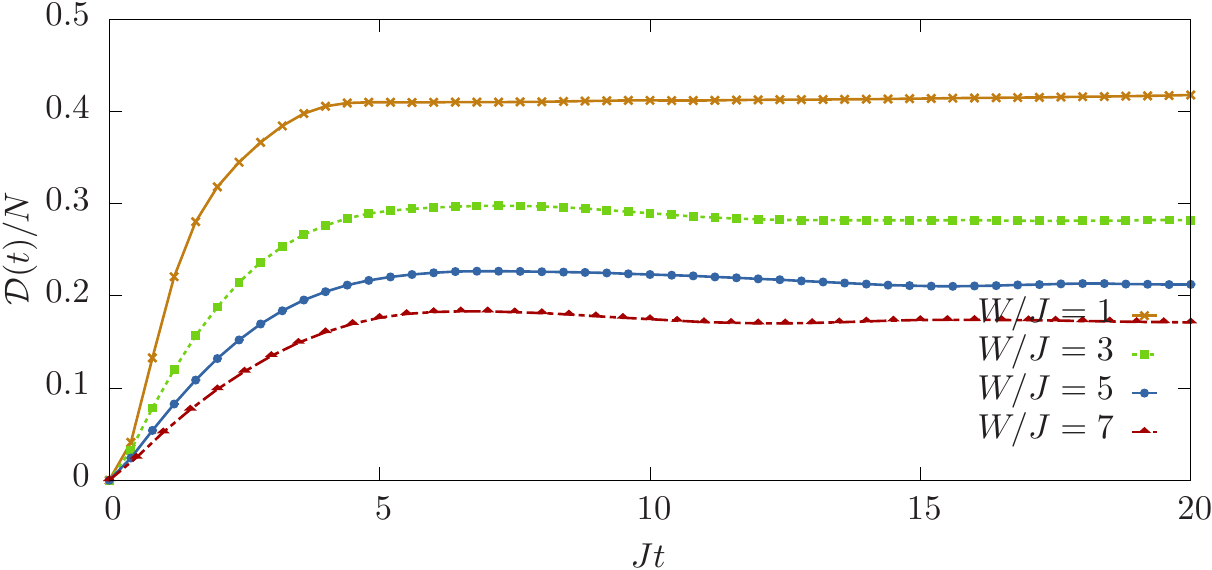}
\caption{(color online) Dynamics of the disorder-averaged Hilbert space distance $\mathcal{D}(t)$ averaged over $800$ disorder realizations for interaction exponent $\alpha=0.5$ , $N=16$, and different disorder strengths.}   
\label{fig:1}
\end{figure}

Since we are interested in localization properties over the entire spectrum, in our numerics we average the results over all initial configurations, which is equivalent to an infinite-temperature initial state. 
Averaging also over disorder, we denote the resulting Hilbert-space distance as $\mathcal{D}(t)$ and the corresponding autocorrelation function as $\chi(t)$. Their relation can then be written as 
\begin{align}
	\mathcal{D}(t) = \frac{N}{2} \left[1-\chi(t)\right] =  \left\langle \frac{1}{2^N} \sum_{\sbold_\initconf} \mathcal{D}_{\sbold_\initconf}(t) \right\rangle_\mathrm{dis}
\label{eq:Daverage}
\end{align}
with $\langle \dots \rangle_\mathrm{dis}$ denoting the disorder average. In Fig.~\ref{fig:1}, the dynamics of the Hilbert-space distance $\mathcal{D}(t)$ is shown for the disordered long-range Ising chain in Eq.~(\ref{eq:defHamiltonianIsing}).

From Eq.~\eqref{eq:Daverage}, it is now straightforward to characterize ergodicity. Since the system can only be ergodic if the spin configuration at large times is uncorrelated with the initial configuration \cite{Anderson1958wx}, we have that $\chi(t\to\infty) = N^{-1} \sum_l \langle \sigma_l^z(t\to\infty) \rangle \langle  \sigma_l^z  \rangle$.
Here, $\langle \dots \rangle$ denotes the average of both the disorder and all initial spin configurations. If the system is ergodic, the long-time value of the local magnetization $\langle  \sigma_l^z(t\to\infty) \rangle$ has to approach its equilibrium value, which in the zero magnetization sector relevant in this work gives $\langle  \sigma_l^z(t\to\infty)  \rangle=0$. Therefore, we find the following necessary criterion for ergodicity:
\begin{align}
	\frac{\mathcal{D}_\infty}{N} = \frac{\mathcal{D}(t\to\infty)}{N} \left\{ 
	\begin{array}{ll} 
	<{1}/{2}\, ,&\,\, \mathrm{nonergodic}\\
	={1}/{2}\, ,&\,\, \mathrm{ergodicity\,\,\,possible}
	\end{array}
 \right.
\label{eq:ergodicityCriterion}
\end{align}
Although $\mathcal{D}_\infty=N/2$ is only a necessary condition for ergodicity (e.g., integrable free fermion models easily satisfy it), we would like to emphasize that the condition $\mathcal{D}_\infty<N/2$ is  \emph{sufficient} for proving nonergodicity because it implies a preservation of a local memory from the initial state. The ergodicity condition for $\mathcal{D}_\infty$ might vary in other cases, e.g., when not working in the zero magnetization sector. Notice that although the localization length is defined for a specific basis (here, we took the most natural choice of configurations in the direction of disorder), to prove nonergodic behavior it is sufficient to demonstrate the criterion $\mathcal{D}_\infty<N/2$ for only one choice of basis.  

As a consequence of relation~\eqref{eq:ergodicityCriterion}, the asymptotic many-body distance $\mathcal{D}_\infty$ behaves fundamentally different from the real-space localization length in a single-particle Anderson insulator. To see this, consider an analogous scenario for a conventional Anderson insulator, and let us again prepare an initially localized wave packet, but now in real space. Evolving the system to infinite time, the mean distance is $\mathcal{D}_\mathrm{AI} \propto \xi$, with $\xi$ the single-particle localization length, as long as we are close to the Anderson transition where the long-distance exponential tails dominate over the nonuniversal short-range contributions. The single-particle localization length $\xi$ is independent of system size $N$ in the localized phase, provided $\xi\ll N$, and diverges when approaching the Anderson transition. The many-body distance $\mathcal{D}_\infty$ on the other hand is always extensive $\mathcal{D}_\infty \propto N$, see Eq.~(\ref{eq:Daverage}), which can be attributed to the unconventionally high connectivity of the underlying graph of spin configurations. Although distances behave differently in the single-particle and many-body case, both allow to detect potential Anderson transitions in real space or Hilbert space, respectively, either via a divergent $\xi$ or via Eq.~(\ref{eq:ergodicityCriterion}).

In practice, and of particular importance for experiments, we can considerably simplify the averaging procedure in Eq.~(\ref{eq:Daverage}), because it is possible to restrict the analysis to one single initial state. For example, one may rotate the local coordinate systems of the spins around the $x$-axis to map $|\sbold_\initconf\rangle$ to the fully polarized state $|{\uparrow\uparrow\dots}\rangle$, i.e., $\sigma_l^z\rightarrow s_l^0\sigma_l^z$. 
Sign flips in Eq.~(\ref{eq:defLocalizationLength_chi}) cancel, but the magnetic fields in Eq.~(\ref{eq:defHamiltonianIsing}) are mapped to $h_l\rightarrow s_l^0 h_l$. If the signs of $h_l$ and $s_l^0$ are uncorrelated, we obtain again an Ising model with random fields. 
Starting from the polarized state has the additional advantage that $\mathcal{D}_{|{\uparrow\uparrow...}\rangle}(t)$ is a simple function of the mean magnetization, i.e., single-site resolved measurements are not necessary. 

\section{Many-body localization in the quantum Ising model with power-law interactions}
\label{sec:results}

We now turn to a detailed analysis of the Hilbert-space distance $\mathcal{D}(t)$. As we will show, based on the ndRG and extensive numerical simulations we find $\mathcal{D}(t)/N<1/2$ for any nonvanishing disorder strength, see Fig.~\ref{fig:2}, indicating that the random Ising model with the considered power-law interactions is MBL. In the following, we will first summarize our main findings in Sec.~\ref{ssec:results}. In Sec.~\ref{ssec:ndRG} we will then discuss how we derived the Hilbert-space distance on the basis of the ndRG, and afterwards provide details about analytical explanations for the absence of ergodicity in the long-ranged regime of $0 \leq \alpha \leq 1$ by analyzing the statistics of resonances in Sec.~\ref{ssec:statistics}. 

\subsection{Results for the many-body localization length}
\label{ssec:results}

\begin{figure}
\centering
\includegraphics[width = \columnwidth]{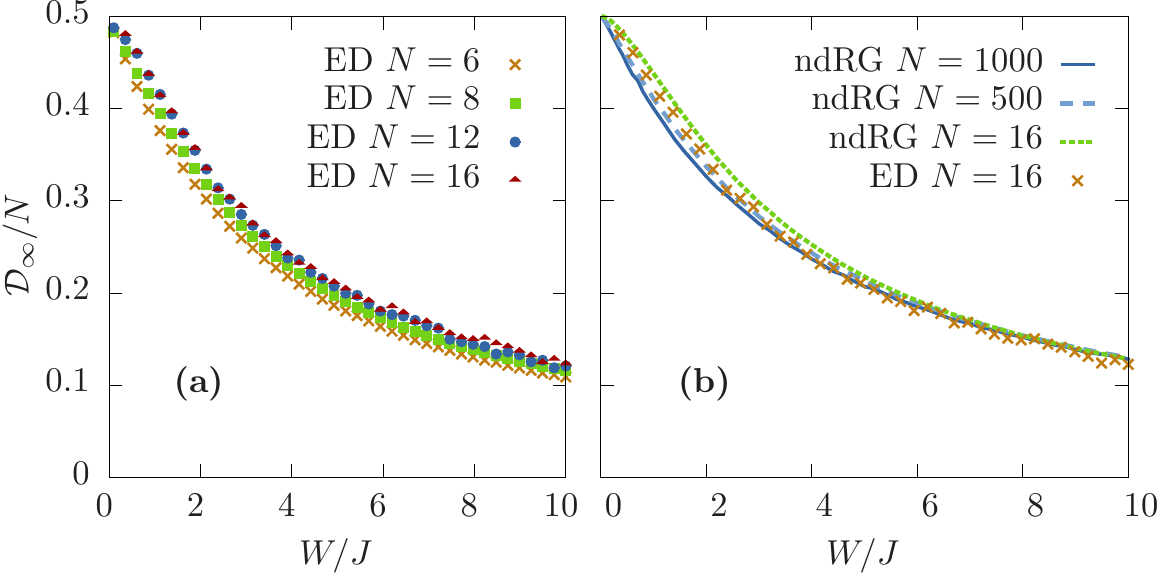}
\caption{(color online) Asymptotic long-time value of the many-body localization length $\mathcal{D}_\infty = \mathcal{D}(t\to\infty)$ for interaction exponent $\alpha=0.5$ as a function of the disorder strength $W/J$. {\bf a:} Results from exact diagonalization (ED) for system sizes $N=6,8,12,16$, averaged over $10^4,200$ disorder realizations.
{\bf b:} $\mathcal{D}_\infty$ from the nonequilibrium dynamical renormalization group (ndRG) for $N=16,500,1000$ ($10^5,10^3,500$ disorder realizations, respectively). Note that the ndRG provides a remarkably good quantitative description over a large range of disorder strengths, see the good match with the ED data at $N=16$ which is included for comparison. The saturation at values $\mathcal{D}_\infty<N/2$ indicates nonergodic behavior.}
\label{fig:2}
\end{figure}

In order to address quantum ergodicity and many-body localization in disordered long-range Ising chains for $\alpha\leq 1$, we use three complementary methods: exact diagonalization, ndRG, as well as an analytical approach on the basis of the statistics of resonances.

In Fig.~\ref{fig:2}, the main results are summarized. In that figure, we show data at $\alpha=0.5$ for the Hilbert-space distance $\mathcal{D}_\infty$ obtained within the ndRG as well as using exact diagonalization. For any nonvanishing disorder strength $W/J>0$, we get $\mathcal{D}_\infty < N/2$. According to the ergodicity criterion in Eq.~(\ref{eq:ergodicityCriterion}) this implies that the system is MBL. Therefore, the autocorrelation function $\chi(t)$ preserves for all times an extensive memory of the initial spin configuration, a behavior that can be attributed to the emergence of local conservation laws in MBL systems \cite{Vosk2012,Huse2013,Serbyn2013a,Ros2014}. 

Let us focus first on the exact diagonalization data in Fig.~\ref{fig:2}a. The simulations have been performed on the basis of a Lanczos algorithm with full reorthogonalization~\cite{Cullum2002}. We have determined $\mathcal{D}_\infty$ by computing the time evolution of $\mathcal{D}(t)$ to large times $Jt \sim \mathcal{O}(10^4)$. Although for increasing system size a tendency towards larger Hilbert-space distances and therefore delocalization is visible, the data for $N=12$ and $N=16$ are already quite close with a weak finite-size dependence at moderate disorder strength, but increasing fluctuations for large disorder. However, for a very weak random field, the finite-size dependence is much stronger. Here, a delocalized phase might still be possible in the thermodynamic limit, although the ndRG discussed in the following gives evidence for a persistence of the MBL phase also in this regime. 

In Fig.~\ref{fig:2}b we show the results obtained using the ndRG, which allows us to consider much larger system sizes up to $N=1000$. Moreover, for $N=16$ we compare ndRG data to exact diagonalization where one can see that the agreement is very good over almost the full range from strong to weak disorder. Deviations between the ndRG, which is constructed for strong disorder, and the ED are very small, especially when bearing in mind that $\mathcal{D}_\infty$ involves a long-time limit of a nonequilibrium quantum real-time evolution, which is a challenging task for perturbative (RG) methods~\cite{Polkovnikov2011kx}. In this light, the very good quantitative description of $\mathcal{D}(t)$ even in the long-time limit underlines the capabilities of the ndRG. Compared to ED, the ndRG can also be used to study very large systems up to $N=1000$ lattice sites. 

The corresponding data is also shown in Fig.~\ref{fig:2}. As one can see, for very large systems the ndRG tends towards localization. In particular, we do not find indications of a strong-coupling divergence which would otherwise point towards the appearance of an ergodic phase. Therefore, the ndRG data gives strong evidence for a MBL phase. This is supported by analytical calculations done in Sec.~\ref{ssec:statistics} where we show on the basis of the statistics of resonances that the system is indeed nonergodic in the regime $0\leq \alpha \leq 1$, independent of disorder strength. Specifically, we find that the many-body eigenstates only occupy a very small fraction of all available states although they are still extended through many-body Hilbert space.

\subsection{Nonequilibrium dynamical renormalization group (ndRG)}
\label{ssec:ndRG}

After having summarized the main results, we now discuss the ndRG implementation for the disordered long-range Ising chain considered in this work. We state here only its main ideas and refer to Appendices~\ref{app:ndRG} and~\ref{app:ndRGMeasure} for methodological details. The ndRG provides an iterative coarse-graining procedure for the full time-evolution operator $\teo(t)$, with the aim to construct an analytically tractable representation of $\teo(t)$ for interacting quantum many-body systems. As we have seen in Figs.~\ref{fig:1} and~\ref{fig:2}, when benchmarking against exact diagonalization, although the ndRG is constructed around the large-disorder limit, its results match remarkably well also in the region $W/J = \mathcal{O}(1)$. 

Starting from the large-disorder limit $W\gg J$, the ndRG eliminates the Ising couplings perturbatively on an iterative basis. Its underlying idea is based on the principle of scale separation: A spin subject to a large magnetic field is energetically decoupled from the remainder of the system. Following this reasoning, we can select the spin in the field with largest magnitude, say at site $\ell$, and remove it from the many-body dynamics by taking its influence on the residual spins into account perturbatively. As explained in Appendix~\ref{app:ndRG}, this leads to renormalized couplings $\mathcal{J}_{mm'}^{\RG}$ and fields $h_m^{\RG}$, which have to be evaluated self-consistently through the RG equations 
\begin{subequations}
\label{eq:RGequations}
\begin{eqnarray}
	h_\ell^{\RG} & =& h_\ell + \sum_{m\not=\ell}  \mathcal{J}_{\ell m} ^2 \frac{2 h_m^{\RG}}{\left(h_m^{\RG}\right)^2 - \left(h_\ell^{\RG}\right)^2} ,\qquad  \label{eq:RGforhl}\\
	h_m^{\RG} & =& h_m -  \mathcal{J}_{\ell m}^2 \frac{2 h_\ell^{\RG}}{\left(h_m^{\RG}\right)^2 - \left(h_\ell^{\RG}\right)^2},\qquad  \label{eq:RGforhm}\\
	\mathcal{J}_{mm'}^{\RG} &  =& \mathcal{J}_{mm'} - \mathcal{J}_{m \ell} \frac{2 h_\ell^{\RG}\sigma_\ell^z}{\left(h_m^{\RG}\right)^2 - \left(h_\ell^{\RG}\right)^2} \mathcal{J}_{\ell m'}.\qquad\label{eq:RGforJ}
\end{eqnarray}
\end{subequations}
These equations bear a strong similarity to a Schrieffer--Wolff transformation to order $(J/W)^2$  with, however, one crucial difference: the right-hand side of the equations involves the \emph{renormalized} magnetic fields. As a consequence, degeneracies with vanishing denominators are lifted, greatly enhancing the stability of the scaling equations. 
Additionally, the Kac prescription, Eq.~\eqref{eq:couplingScaling}, \cite{Kac1963} ensures the extensivity of the spin-interaction contribution to the total energy, rendering the ndRG well-controlled even in the case of long-range interactions.  In particular, we do not find any indications of a strong-coupling divergence, which indicates that the ndRG is well-controlled in the present scenario.

The RG equation \eqref{eq:RGforJ} for the couplings additionally involves the spin projection $\sigma_\ell^z$ of the eliminated spin. This projection, after the RG step, is a constant of motion, so we can treat it as a $c$-number. As long as we are in a nonergodic phase and the system retains a memory of the initial state, we can replace $\sigma_\ell^z \to \langle \sigma_\ell^z\rangle$ by its initial value, up to corrections that are of the order $(J/W)^2$. This means that within this prescription we can a priori only describe the MBL phase of the model. However, the breakdown of the ndRG could potentially also reveal an approach towards the MBL transition. 
Note also that the renormalization introduces a randomness in the couplings. The initial restriction to a randomness in the field terms is therefore not a crucial ingredient of the considered model.

Storing the field of the removed spin $h_\ell^{\RG}$ into memory, calling it $h_\ell^{\ast}$, we remove it from the dynamics. We then repeat the RG step defined by Eq.~\eqref{eq:RGequations}, choosing the next spin with the (renormalized) field of largest magnitude. 
By successively eliminating all spins, the ndRG scheme prescripes a unitary transformation $U$ [see Eq.~\eqref{eq:teoAfterRG}] to a renormalized model, where all couplings between spins are removed, $\teo(t)= U^\dag \teo_\ast(t) U$, with $\teo_\ast(t) = \ue^{-i H_0^\ast t}$ and $H_0^\ast=\sum_{i} h_i^\ast \sigma_i^z$. 

The simple form of the resulting renormalized Hamiltonian allows for the calculation of the autocorrelation function defined  in Eq.~\eqref{eq:defLocalizationLength_chi}, as explained in detail in Appendix~\ref{app:ndRGMeasure}. 
Using a recently introduced scheme \cite{Heyl2014} for evaluating expectation values of local observables within techniques such as the ndRG, one obtains for a single disorder realization, up to second order in the renormalized coupling strengths $\mathcal{J}_{lm}^\RG$, 
\be
	\chi_{\sbold_0}^\infty = \frac{1}{N}\sum_{l} \exp\left\{\sum_{m}\frac{4\left({\mathcal{J}}_{lm}^\RG\right)^2\left({h}_l^\RG s_l^0 - {h}_m^\RG s_m^0\right)^2}{-\left[\left({h}_l^\RG\right)^2-\left({h}_{m}^\RG\right)^2\right]^2} \right\},
\ee
where $\chi_{\sbold_0}^\infty\equiv \chi_{\sbold_0}(t\to\infty)$, and ${h}_l^\RG$ are the renormalized fields at the step where the coupling ${\mathcal{J}}_{lm}^\RG$ is removed.

We numerically performed the ndRG to calculate $\mathcal{D}_\infty=\mathcal{D}(t\to \infty)$. As already summarized in Sec.~\ref{ssec:results}, Figure~\ref{fig:2} displays the results for one representative example $\alpha=0.5$, but other values of $\alpha \leq 1$ give qualitatively similar outcomes. This ndRG data compares remarkably well with ED, which gives additional confidence in the validity of the ndRG approach. We attribute the reliability of the ndRG to the structure of the RG equations in Eqs.~(\ref{eq:RGforhl}-\ref{eq:RGforJ}) which relies on a \emph{self-consistent} determination of the renormalized Hamiltonian parameters. In particular, potential resonances with small energy denominators are lifted which leads to a substantial increase in stability of the RG equations.
The ndRG assumes that there are not too many such resonances. As we will see in the next Section, this is a well-justified assumption in the range $0\leq \alpha\leq 1$ for all values of disorder.

\subsection{Statistics of resonances}
\label{ssec:statistics}

Our ED and ndRG studies showed nonergodic behavior of the model~(\ref{eq:defHamiltonianIsing}). We will now explain this finding analytically for the parameter regime $0\leq \alpha \leq 1$ via the statistics of resonances, which has proven valuable in the context of single-particle localization phenomena \cite{Anderson1958wx,Abou-Chacra1973bl,Nguyen1985,Medina1992} and has recently been extended to the interacting many-body context \cite{Altshuler1997hx,DeLuca2014,Laumann2014,Ros2014}. This will allow us to characterize localization and ergodicity in the lattice of spin configurations.  For the moment, let us first concentrate on $\alpha=0$ where it has already been shown that the system is many-body localized~\cite{Ioffe2010}. The results obtained for this limit will also allow us to establish a many-body localized phase in the entire regime $0\leq \alpha \leq 1$.

For $\mathcal{J}>0$, the eigenstates $|\tilde{\sbold}\rangle = \ue^S |\sbold\rangle$ are perturbatively connected to the $\mathcal{J}=0$ eigenstates via a unitary transformation, whose generator $S$ can be obtained through a Schrieffer--Wolff transformation \cite{Schrieffer1966xr}. 
To lowest order in $\mathcal{J}/W$, the amplitude $A_\Lambda$ connecting two configurations along a given trajectory reads (see Appendix \ref{app:statisticsResonances}) 
\be
	A_\Lambda = \prod_{\nu=1}^\Lambda \frac{\mathcal{J}/2}{\Delta \varepsilon_\nu}.
	\label{eq:ALambda}
\ee
The details about the trajectory enter through the energy differences $\Delta\varepsilon_\nu=E_\sbold-E_{\sboldbar}$ between `neighboring' configurations $|\sbold\rangle$ and $|\sboldbar\rangle$, i.e., configurations connected by the Ising interaction with $V_{\sbold,\sboldbar}\neq 0$ in Eq.~\eqref{eq:AndersonModel}. The Ising interaction involves spin flips at two real-space sites $l$ and $m$.  
Thus, we have $\Delta \varepsilon_\nu = \pm h_l \pm h_m$, with signs depending on whether the spins are flipped from $\ua$ to $\da$ or vice versa.  For uniformly distributed magnetic fields, $h_l\in[-W,W]$, the probability distribution for the magnitude of nearest-neighbor amplitudes $|A_1|$ is
\be
	P(|A_1|) = \frac{2}{Z^2 |A_1|^3}\left[Z|A_1|-1\right],\quad Z=\frac{4W}{\mathcal{J}}, 
	\label{eq:PA1}
\ee
for $|A_1|>Z^{-1}$ and zero otherwise, see also Ref.~\onlinecite{Altshuler1997hx}. The fraction of nearest-neighbor states which share a resonance is given by the probability $P_1(C) = (ZC)^{-1} [2-(ZC)^{-1}]$ that the amplitude $|A_1|$ exceeds a given value $C$, with $C$ setting the threshold for resonances. Since $\mathcal{J}=J/N$, we have $Z=4NW/J$, implying $P_1(C)\to 0$ for any fixed $C>0$ when $N\to \infty$ which is independent of the precise threshold value $C$. How this property relates to ergodicity will be discussed below.

Although resonances between close generations are extremely sparse, we will now show that the eigenstates are nevertheless highly extended. For multiple hopping processes, the situation is in general much more difficult than for the hopping to the next generation analyzed above. It is, however, possible to simplify the analysis substantially through a controlled approximate mapping onto a much simpler subgraph that can be solved analytically as we will show now. 
Consider a first hopping process from a ``root'' site $|\sbold_0\rangle$ to one site $|\sbold\rangle$ of generation $1$. From this particular $|\sbold\rangle$, there is one path back to the root, there are $2(N-2)$ paths to the same generation, and  $\binom{N-2}{2} = (N-2)(N-3)/2$ trajectories to the next generation. In the thermodynamic limit $N\to \infty$, it is therefore possible to only consider the latter paths, see also Ref.~\onlinecite{Altshuler1997hx}. 
Extending the same argument to trajectories with $\Lambda>2$, one obtains an effective directed graph including only those trajectories that minimize the length between the connected sites. Such subgraphs are also known in the context of the forward-scattering approximation~\cite{Nguyen1985,Medina1992,Laumann2014}, which is controlled by the perturbation strength $J/W$~\cite{Nguyen1985,Medina1992}. It is important to emphasize that in the present context the mapping onto the subgraph is additionally controlled by the large connectivity of the underlying Hilbert space graph towards higher generations, similar to Ref.~\cite{Altshuler1997hx}. As we will discuss below, for $\alpha>1$, this mapping is not well-controlled, restricting the use of this graph to $\alpha\leq 1$. 
For $\Lambda\to N/2$ the connectivity towards higher generations becomes smaller, decreasing the accuracy of the description via the subgraph. In this case, as in the forward-scattering approximation, the mapping still remains well-controlled due to the perturbation strength. 
Notice that the graph used here is different from a Bethe lattice as used in Ref.~\onlinecite{Altshuler1997hx}. 

On this reduced subgraph, it is possible to use a saddle-point approximation for $\Lambda \gg 1$ in order to determine the probability $P_\Lambda(C)$ that the magnitude of the amplitude $A_\Lambda$ exceeds a given $C$, yielding (see Appendix \ref{app:statisticsResonances})  
\be
	\label{eq:PSingleTrajectory}
	P_\Lambda(C) = \frac{1}{\sqrt{2\pi\Lambda}} \frac{1}{C  \log(Z)} \left[ \frac{2\ue\log(Z)}{Z} \right]^\Lambda,
\ee
for $Z^{-\Lambda}\ll C<1$. The probability $\overline{P}_\Lambda$ that none of the trajectories has an amplitude larger than $C$, see Ref.~\onlinecite{Altshuler1997hx}, is $\overline{P}_\Lambda(C)=[1-P_\Lambda(C)]^{n_\Lambda} \approx \exp[-n_\Lambda P_\Lambda(C)]$, where $n_\Lambda = 2^{-\Lambda} N!/(N-2\Lambda)!$ is the number of trajectories connecting the root to sites in generation $\Lambda$. Using Stirling's approximation, we obtain $n_\Lambda\to [K(\lambda)]^\Lambda$ for $\Lambda,N \gg 1$  with $\lambda = \Lambda/N$ and $K(\lambda) = N^2 (1-2\lambda)^{2-1/\lambda}/e^2$. Because $n_\Lambda P_\Lambda(C) \propto [2\ue \log(Z) K(\lambda)/Z]^\Lambda \propto [N\log(N)]^\Lambda$, we have $\overline{P}_\Lambda(C) \to 0$ for $N\to\infty$, i.e., there is at least one trajectory and therefore one site in generation $\Lambda$ that is strongly connected to the root. With probability $1$ each eigenstate extends to arbitrary distance, but restricted to a small fraction of the available states \cite{Altshuler1997hx,DeLuca2014}, since $P_1(C)$ vanishes in the thermodynamic limit.

We now generalize this analysis to the case $1\geq\alpha>0$. The couplings $\mathcal{J}_\nu = \mathcal{J}_{lm}$ appearing in the amplitudes $A_\Lambda = \prod_\nu \mathcal{J}_\nu/2 \Delta \varepsilon_\nu$ now depend explicitly on the specific spins that are flipped along the trajectory. However, to show that the system is still nonergodic, it suffices to consider an upper bound for $|A_\Lambda|$, obtained by replacing $\mathcal{J}_{lm}\to \mathcal{J}$ by its nearest-neighbor value. Following the same steps as above, this implies $P_1(C)\to 0$ for $N\to \infty$, because $\mathcal{J}$ decays as $J/N^{1-\alpha}$ for $0<\alpha<1$ and as $J/\log(N)$ at $\alpha=1$. 

Summarizing, the statistics of resonances reveals the structure of the eigenstates in the disordered long-range Ising model and therefore its ergodicity properties. 
For $0\leq \alpha\leq 1$, resonances between nearest-neighboring spin configurations are vanishingly sparse in the thermodynamic limit because $P_1(C)\to0$ for $N\to\infty$. 
Hence, eigenstates occupy only a vanishing fraction of Hilbert space. Therefore, they are nonergodic. But, remarkably, eigenstates are still extended \cite{Altshuler1997hx,DeLuca2014} as there is always at least one resonant trajectory connecting a root configuration to one site in generation $\Lambda$ with $\Lambda \gg 1$. Interestingly, localization in many-body \emph{Hilbert} space is possible although single-particle excitations can delocalize in \emph{real} space for sufficiently long-ranged interactions \cite{Burin2006,Burin2014}. In our model, we do not find indications for a phase where eigenstates are nonergodic and also localized as has been observed for Cayley trees~\cite{Altshuler1997hx}. We attribute this to the particular relation between the connectivity $K$ of our lattice, $K \propto N^2$, and the effective disorder strength $Z\propto N$. Thus, we always have that $K \gg Z$, a regime which does not allow for states which are both nonergodic \emph{and} localized~\cite{Altshuler1997hx}.

The situation is more complex for $\alpha>1$, where, contrary to $\alpha \leq 1$, the couplings $\mathcal{J}$ do not decay as a function of system size $N$. This scaling, however, is crucial for the above analysis, preventing the use of the same methods. In particular, the applicability of the forward scattering approximation becomes much less controlled in this case because higher-order processes can dominate over lower-order processes  as we will discuss now. Consider, for example, the coupling of a spin configuration to a configuration in generation 1, where the two flipped spins in real space are at a distance of $r$. The corresponding coupling amplitude is $A_1=\mathcal{J}/\Delta \varepsilon\, r^\alpha$. The same configurations can also be coupled by second-order processes, for example, one where first the spins at position $0$ and $r-d$ are flipped, followed by a second hopping within generation 1 involving the spins at $r-d$ and $r$. The corresponding amplitude is $A_2 = \mathcal{J}^2/\Delta \varepsilon_1 \Delta \varepsilon_2\, (r-d)^\alpha d^\alpha$, with $\Delta \varepsilon_{1,2}$ the energy difference of the first and second spin flip processes. Let us consider the possibility that the second-order process dominates, i.e., $A_2>A_1$. This leads to the condition $r/d<c d/(cd-1)$ for $r>d>1$, with $c=[{\mathcal{N}(\alpha)}\Delta \varepsilon_1\Delta \varepsilon_2/J\Delta \varepsilon]^{1/\alpha}$. For $\alpha>1$, this condition can always be fulfilled whereas for $\alpha\leq1$ this is not the case, because then $c\to\infty$ for $N\to\infty$. More precisely, for $\alpha\leq 1$ one obtains that $r/d<c d/(cd-1)\to 1$, resulting in a contradiction with $r>d>1$. In other words, for $\alpha\leq 1$ second-order processes within generation 1 can be safely neglected in the thermodynamic limit. This is a further justification for the applicability of the forward scattering approximation and the above use of the reduced subgraph. For $\alpha>1$ instead, taking only the probability for first-order resonances as a criterion for ergodicity requires care. However, it is still possible to address the delocalization of single-particle excitations in real space on the basis of the first-order resonances~\cite{Burin2006}.

\section{Conclusions}
\label{sec:conclusions}

In this article, we have studied many-body localization in Ising models with slowly decaying power-law interactions in a disordered transverse field, which are relevant for experiments with polar molecules, Rydberg atoms, and trapped ions. We have presented numerical and analytical calculations predicting an infinite-temperature many-body localized phase. Consequently, these systems show nonergodic behavior throughout the entire spectrum. 

Moreover, in Eq.~(\ref{eq:defLocalizationLength}), we have introduced an experimentally accessible observable that quantifies distances in Hilbert space. It can be seen as the analog of the Anderson localization length in the many-body context  and thus allows one to experimentally access fundamental properties of many-body localized phases. A straightforward sequence to measure it in a spin system would be: (i) initialize all spins in the $\uparrow$ state; 
(ii) time evolve under one realization of the disordered model; (iii) measure the mean magnetization; 
and (iv) average the results over disorder realizations. This sequence is general and can be exploited in other experimental contexts, simply by inserting in the time evolution (ii) the appropriate disordered many-body model.


\begin{acknowledgments}
We acknowledge helpful discussions with A.~Gorshkov, C.~Monroe, F.~Pollmann, P.~Richerme, Wu Yukai, and P.~Zoller.   
This work was supported by the Deutsche Akademie der Naturforscher Leopoldina (grant No.~LPDS 2013-07), EU IP SIQS, SFB FoQuS (FWF Project No.~F4016-N23), and ERC synergy grant UQUAM. The ED algorithm uses the Armadillo linear algebra libraries \cite{Sanderson2010}.

\end{acknowledgments}


\appendix

\section{Statistics of resonances}
\label{app:statisticsResonances} 

Ergodicity, i.e., delocalization in many-body Hilbert space, is driven by the proliferation of resonances between sites in Hilbert space~\cite{Altshuler1997hx}. In this section, we provide technical details about the statistics of resonances presented in the main text. In the context of localization phenomena, similar analyses have proven very valuable both for Anderson~\cite{Anderson1958wx,Abou-Chacra1973bl,Nguyen1985,Medina1992} and many-body localization~\cite{Altshuler1997hx,DeLuca2014,Laumann2014,Ros2014}. 

Our starting point is the Hilbert-space lattice defined by the spin configurations $|\sbold \rangle = |s_1,\dots,s_N\rangle$ with $s_l=\left|\ua\right\rangle,\left|\da\right\rangle$. Without the Ising coupling between the spins ($\mathcal{J}=0$), the Hamiltonian Eq.~(1) of the main text becomes purely local and its eigenstates are the spin configurations $|\sbold \rangle$. 
Within standard perturbation theory, for nonzero spin interactions ($\mathcal{J}>0$), the lowest-order correction to the eigenstates connects states with Hamming distance $2$ that can be reached by flipping two spins via the interaction, i.e., states $|\sbold\rangle$ and $|\sboldbar\rangle$ where $V_{\sbold,\sboldbar}\neq 0$ in Eq.~(2). For the statistics of resonances as given in main-text Eq.~(8), however, we are also interested in states separated by large Hamming distance, which is far beyond low-order perturbation theory. In the following, we provide a general scheme for determining amplitudes for far distant spin configurations in the disordered long-range Ising chain. 

For $\mathcal{J}>0$, the Ising Hamiltonian $H_\mathrm{Ising}$ can be diagonalized approximately using a Schrieffer--Wolff transformation~\cite{Schrieffer1966xr}, 
\be
	\ue^{-S} H_\mathrm{Ising} \ue^{S} = \sum_l h_l \sigma_l^z + \mathcal{O}(\mathcal{J}^2/W) = H_0\,,
\ee
up to perturbative corrections of the order $\mathcal{J}^2/W$. The generator $S$ of the transformation is chosen such that $[H_0,S]=\sum_{l<m} \mathcal{J}_{lm}\sigma_l^x \sigma_m^x$, which is achieved by 
\be
	S = \sum_{l<m} S_{lm},
\ee
with
\begin{align}
	S_{lm} = &  i \frac{\mathcal{J}_{lm}}{4} \left[\frac{1}{h_m+h_l} \left( \sigma_l^x \sigma_m^y + \sigma_l^y \sigma_m^x \right) + \right. \nonumber \\
	& \left. +  \frac{1}{h_m-h_l} \left( \sigma_l^x \sigma_m^y - \sigma_l^y \sigma_m^x \right)\right].
\end{align}
If $|\sbold\rangle$ is an eigenstate of $H_0$, then $|\tilde{\sbold}\rangle = \ue^{S}|\sbold\rangle$ is an approximate eigenstate of $H_\mathrm{Ising}$. Expanding the exponential $\ue^S$, we get $|\sbold\rangle=\sum_{n=0}^\infty S^n |\sbold\rangle/n!$, where $S^n$ contains all contributions of the order $(\mathcal{J}/W)^n$. For a given $n$, $S^n|\sbold\rangle$ can be decomposed into individual trajectories connecting the configuration $|\sbold\rangle$ to other configurations. 
A specific trajectory reaching state $|\sbold'\rangle$ will have the amplitude
\be
	A_\Lambda= \ue^{i\varphi_\Lambda}\prod_{\nu=1} ^\Lambda \frac{\mathcal{J}_\nu/2}{\Delta \varepsilon_\nu}. 
\ee
The details of the particular trajectory are contained in the combined index $\nu=(l,m)$, which keeps track of the spins in real space that have been flipped on the trajectory, with $\mathcal{J}_\nu = \mathcal{J}_{lm}$ and $ \Delta\varepsilon_\nu = \pm h_l \pm h_m$. The signs in $\Delta \varepsilon_\nu$ depend on whether the spins on sites $l$ and $m$ have been flipped from $\ua$ to $\da$ or vice versa. These signs, as well as the overall phase $\varphi_\Lambda$, however, will not be important for what follows, because we will only be interested in the magnitude of the objects $A_\Lambda$. 

Focusing first on the case $\alpha=0$, we have that $\mathcal{J}_\nu =\mathcal{J}$, and randomness enters only via the energy denominators, i.e., for the statistics of the amplitudes $A_\Lambda$ we need the distribution $D_\Lambda$ of the denominators, 
\be
	D_\Lambda = \prod_{\nu=1}^\Lambda \frac{W}{\Delta \varepsilon_\nu}. 
\ee
Let us first consider hopping processes between nearest-neighboring Hilbert-space sites, i.e., $\Lambda=1$. For a uniform distribution of $h_l\in[-W,W]$, the probability distribution $P(|D_1|)$ for the absolute value $ |D_1|$ can be calculated straightforwardly, yielding $P(|D_1|) = (2|D_1|-1)/(2 |D_1|^3)$. For the full amplitudes $A_1=D_1 \mathcal{J}/(2W)$, one obtains
\be
	P(|A_1|) = \frac{2}{Z^2 |A_1|^3} \left[ Z|A_1| - 1\right], \qquad Z = \frac{4W}{\mathcal{J}}, 
\ee
the result quoted in Eq.~(6) of the main text. For the derivation, see also Ref.~\onlinecite{Altshuler1997hx}.

Importantly, for the considered trajectories in the derived effective graph of the main text, which include only hopping processes that increase the generation, the energy denominators are independent random variables. Let us introduce the new variables $x_\nu = \log(W/\Delta \varepsilon_\nu)$. The probability distribution $p_\Lambda(X)$ for $X=\sum_\nu x_\nu$ (i.e., $\ue^X\equiv A_\Lambda$) is then obtained via Fourier transformation, 
\begin{align}
	p_\Lambda(X) = & \int dx_1 \dots dx_\Lambda p_1(x_1) \dots  p_1(x_\Lambda) \delta(X-\sum_\nu x_\nu) \nonumber \\
	& = \frac{1}{2\pi} \int d\mu e^{i \mu X} \left[ \int dx p_1(x) \ue^{-i\mu x} \right]^\Lambda, 
\end{align}
with $p_1(x) = (2\ue^{-x}-\ue^{-2x})/2$. We get for the Fourier transform $p_1(\mu) = \int dx p_1(x) \ue^{-i\mu x} =  2^{i\mu+1}[ (1+i\mu)^{-1} - (2+i\mu)^{-2} ]$. For $\Lambda \gg 1$, one can use a saddle-point approximation to obtain $p_\Lambda(X)$. The saddle point $\mu^\ast$ occurs at
\be
	\mu^\ast = i\frac{1}{2\overline{x}} \left[2-3\overline{x} + \sqrt{\overline{x}^2 +4} \right], \qquad \overline{x} = \frac{X}{\Lambda} + \log(2)>2.
\ee
Performing the saddle-point integral, one obtains 
\begin{align}
	p_\Lambda(X) = & \frac{1}{\sqrt{2\pi n} } \frac{1}{\sqrt{\overline{x}^2 + 4 - 2\sqrt{\overline{x}^2 + 4} }} \times \nonumber \\ & \times \left[ \frac{2\ue\overline{x}^2}{2 + \sqrt{\overline{x}^2 +4}} \ue^{-\frac{3}{2} \overline{x} + \frac{1}{2} \sqrt{\overline{x}^2 + 4}}\right]^\Lambda.
\end{align}
From this expression, one can obtain the distribution $P(|A_\Lambda|)$ of the full amplitude.  For amplitudes $Z^{-\Lambda} \ll |A_\Lambda| <1$, where  $|\log(|A_\Lambda|)| \ll |\log(Z^n)|$---precisely those that characterize the resonances---the distribution $P(|A_\Lambda|)$ becomes 
\be
	P(|A_\Lambda|) = \frac{1}{\sqrt{2\pi\Lambda}} \frac{1}{|A_\Lambda|^2  \log(Z)} \left[ \frac{2\ue\log(Z)}{Z} \right]^\Lambda.
\ee
Integrating from some constant $C$ to infinity gives the probability used in Eq.~(7) of the main text.

\section{Nonequilibrium dynamical renormalization group (ndRG)}
\label{app:ndRG}

In this Appendix, we present methodological details for the nonequilibrium dynamical renormalization group (ndRG)~\cite{Heyl2013}. The ndRG provides a coarse-graining procedure that establishes an analytically tractable representation of the full time-evolution operator 
\be
	\label{eq:fullTimeEvolutionOperator}
	\teo(t) =  \mathcal{T} \ue^{-i \int_0^t dt' \,\, H(t')}
\ee
of complicated many-body problems. This is achieved by successively eliminating high-energy contributions, thereby generating an effective theory for the low-energy degrees of freedom. In Eq.~\eqref{eq:fullTimeEvolutionOperator}, $H(t)$ denotes the potentially time-dependent Hamiltonian of the system, and $\mathcal{T}$ is the time ordering prescription. 

In the present case, the system is initially prepared in a specific spin configuration $|\sbold\rangle$, 
and we are interested in the time evolution with the disordered long-range Ising Hamiltonian 
\be
	H(t>0)=H_\mathrm{Ising}=H_0+V\,,
\ee
where $H_0=\sum_{i} h_i \sigma_i^z$, and $V = \sum_{i<j} \mathcal{J}_{ij} \sigma_i^x \sigma_j^x$. 
The associated time-evolution operator is
$
	\teo(t) = \exp({-i H_\mathrm{Ising} t})
$.

In the limit of strong disorder $W\gg J$, the largest energy scale will be the magnetic field of largest magnitude, located, say, at spin $\ell$. It is then convenient to separate the interactions that involve this spin, denoted by $V_\ell$, from those which do not, $\overline{V}_\ell$:
\be
	V = V_\ell + \overline{V}_\ell \qquad V_\ell = \sigma_\ell^x \sum_{m} \mathcal{J}_{\ell m} \sigma_m^x, \qquad \overline{V}_\ell = V - V_\ell,
\ee
Following Ref.~\onlinecite{Heyl2013}, the ndRG removes all couplings involving the spin $\ell$ using a unitary transformation $e^{S^{(\ell)}}$ on the full time-evolution operator, $\teo(t) = \ue^{-S^{(\ell)}}  \teo^{(\ell)}(t) \ue^{S^{(\ell)}}$. This transformation yields a renormalized model
\be
	\teo^{(\ell)}(t) = \ue^{-i H_\mathrm{Ising}^{(\ell)} t}.
\ee
(Throughout this Appendix, we use a subindex to denote the site of the spin $\ell$ and a bracketed superindex to denote the RG step where spin $\ell$ is integrated out.)  
To second-order accuracy, i.e., including terms up to $\mathcal{O}[(J/W)^2]$, the renormalized Hamiltonian  $H_\mathrm{Ising}^{(\ell)}$ after this RG step is given by
\be
	\label{eq:HIsingrenorm}
	H_\mathrm{Ising}^{(\ell)}= H_0^{(\ell)} + V^{(\ell)} = H_0 + \overline{V}_\ell + \frac{1}{2}\left[ S^{(\ell)},  V_\ell\right].
\ee
Here, $H_0^{(\ell)}=\sum_m h_m^{(\ell)} \sigma_m^z$ denotes the renormalized free part of the Hamiltonian with renormalized magnetic fields $h_m^{(\ell)}$, and $V^{(\ell)}$ are the renormalized spin interactions that remain after eliminating spin $\ell$. 

The generator $S^{(\ell)}$ of the unitary transformation is determined by the equation
\be
S^{(\ell)}(t)-S^{(\ell)} = i \int_0^t dt'\,\,\, V_\ell (t')\,, 
\ee
where $V_\ell(t) = e^{i H_0^{(\ell)}t} V_\ell e^{-i H_0^{(\ell)}t}$, and $S^{(\ell)}(t) = e^{i H_0^{(\ell)}t}S^{(\ell)} e^{-i H_0^{(\ell)}t}$.  
This gives $S^{(\ell)}=\sum_{m\neq\ell} S^{(\ell)}_m$, with 
\be
S^{(\ell)}_m = i \frac{\mathcal{J}_{\ell m}}{4} \left(\frac{ \sigma_\ell^x \sigma_m^y + \sigma_\ell^y \sigma_m^x }{h_m^{(\ell)}+h_\ell^{(\ell)}}  +  \frac{\sigma_\ell^x \sigma_m^y - \sigma_\ell^y \sigma_m^x }{h_m^{(\ell)}-h_\ell^{(\ell)}} \right)\,. 
\label{eq:generator}
\ee 
Using this transformation in Eq.~\eqref{eq:HIsingrenorm}, one obtains the RG equations for the fields and couplings given in Eq.~\eqref{eq:RGequations}. 

We can now repeat the above RG scheme for the spin with second largest magnetic field amplitude (after renormalization), and proceed this way from high to low energies. 
Iteratively eliminating all couplings, one obtains an analytically tractable representation of the time evolution operator, 
\be
	\label{eq:teoAfterRG}
	\teo(t) = U^\dag \teo_\ast(t) U\,, \quad U= \mathcal{T}^{(\ell)} \exp\left(\sum_\ell S^{(\ell)}\right)
\ee
where $\teo_\ast(t) = \ue^{-i H_0^\ast t}$ is diagonal. 
The prescription $\mathcal{T}^{(\ell)}$ denotes energy ordering, with $S^{(\ell)}$ to the right for spins in larger fields, analogous to common time ordering. 

Note that the validity of the scale separation underlying this ndRG algorithm assumes that there are few resonances, i.e., the dominant energy scale is $h_l$ rather than $J/|h_l-h_m|$. Close to a transition to ergodicity, such resonances proliferate, which would lead to a breakdown of the ndRG procedure. As the stability of the numerical algorithm demonstrates, in the parameter ranges considered in this article the influence of such resonances is small, and ndRG always predicts a nonergodic behavior.

\section{Hilbert-space distance using ndRG}
\label{app:ndRGMeasure}

Within the above ndRG procedure, observables can conveniently be evaluated by the scheme introduced in Ref.~\onlinecite{Heyl2014}, which we outline now for the specific case of the two-time correlator as defined in Eq.~\eqref{eq:defLocalizationLength_chi}, $\chi_{\sbold_\initconf}(t) = {N}^{-1} \sum_{m} \left\langle \sbold_\initconf| \sigma_m^z(t)\sigma_m^z |\sbold_\initconf \right\rangle \equiv {N}^{-1} \sum_{m} \chi_{\sbold_\initconf,m}(t)$. 

Using the renormalized time-evolution operator, Eq.~\eqref{eq:teoAfterRG}, we can write 
\be
\label{eq:chil}
\chi_{\sbold_\initconf,m}(t) = \left\langle \sbold_\initconf| \, U^\dag \teo_\ast^\dag(t) U\,\,\sigma_m^z\,\, U^\dag \teo_\ast(t) U\,\, \sigma_m^z |\sbold_\initconf \right\rangle.\quad
\ee
To evaluate this equation, it is convenient to perturbatively eliminate the energy-ordering prescription $\mathcal{T}^{(\ell)}$ appearing in $U$ by using a Magnus expansion:
\be
	U = \exp\left( \sum_\ell S^{(\ell)} + \mathcal{O} [(J/W)^2] \right) \approx  \exp\left( \sum_\ell S^{(\ell)} \right).
\ee
In principle, for consistency, terms of order $(J/W)^2$ should be taken into account. In the expectation value $\left\langle\sbold_\initconf| \sigma_m^z(t)\sigma_m^z |\sbold_\initconf \right\rangle$, however, they can yield finite contributions only if two terms of order $(J/W)^2$ collaborate, so that the corresponding correction to the final result is of the order $(J/W)^4$. This is beyond the desired accuracy of the present calculation. Notice that this argument only holds for initial states that are eigenstates of $H_0$. 

The summation over $S^{(\ell)}$ involves any given spin $m$ several times: it gets repeatedly renormalized by $S^{(\ell)}_m$ [see Eq.~\eqref{eq:RGequations}] until it becomes the spin in the field with largest magnitude. It is then removed from the many-body dynamics and obtains a final renormalization from all remaining spins via $S^{(m)}$. 
For the following, it will, therefore, be useful to split $\sum_{\ell}S^{(\ell)}$ into the part that contains the spin $m$, which we denote 
as $S_m$, 
and the part that does not, $\overline{S_m}=\sum_{\ell}S^{(\ell)}-S_m$. 

Using the Baker--Campbell--Hausdorff formula, these two contributions can be separated, yielding
\be
U = U_m\, \overline{U_m}\,,\quad U_m\equiv \exp\left(S_m\right),\quad \overline{U_m}\equiv \exp\left(\overline{S_m}\right) \,,
\ee
up to corrections that again only contribute to the final result for the correlation function $\chi_{\sbold_0,m}$ to order $(J/W)^4$. Due to the special structure of the transformation described by Eq.~\eqref{eq:generator}, we have $\sigma_m^z U_m =U_m^\dag \sigma_m^z$, \cite{Heyl2014} whereas $\sigma_m^z\overline{U_m}=\overline{U_m} \sigma_m^z$. 
Inserting these relationships into Eq.~\eqref{eq:chil} and commuting the first $\sigma_m^z$ to the right, we obtain 
\be
\label{eq:chil2}
\chi_{\sbold_\initconf,m}(t) = \left\langle \sbold_\initconf^{\phantom{\dag}}\right| \, 
U_m^\dag \overline{U_m}^\dag  \teo_\ast^\dag(t) (U_m)^2 \teo_\ast(t) U_m^\dag \overline{U_m} 
 \left|\sbold_\initconf^{\phantom{\dag}} \right\rangle\,.
\ee
In this expectation value, terms of the type $\left[\overline{U_m}^\dag,{U_m}^\dag\right]$ contribute only if they appear pair-wise, so neglecting them gives again corrections only of order $(J/W)^4$. Thus, we can commute $\overline{U_m}^\dag$ through to the right to annihilate it, and write 
\bea
\label{eq:chil3}
\chi_{\sbold_\initconf,m}(t) &=& \left\langle \sbold_\initconf^{\phantom{\dag}}\right| \, 
  U_m^\dag U_m(t)^2 U_m^\dag
 \left|\sbold_\initconf^{\phantom{\dag}} \right\rangle \\
 &=& \left\langle \sbold_\initconf^{\phantom{\dag}}\right| \, 
  \ue^{-S_m} \ue^{2 S_m(t)} \ue^{- S_m }
 \left|\sbold_\initconf^{\phantom{\dag}} \right\rangle \,.
\eea
Here, we understand $U_m(t)= \teo_\ast^\dag(t) U_m \teo_\ast(t)$ as the time evolution under the renormalized Hamiltonian $H_0^\ast$, which, being diagonal, is easily computed following Heisenberg's equations of motion. 

Finally, a cumulant expansion allows evaluating the expectation values, 
\begin{subequations}
\bea
\label{eq:chil4}
\chi_{\sbold_\initconf,m}(t) &\approx& \exp\left\{ 
\left\langle \sbold_\initconf^{\phantom{\dag}}\right| \, 
2 \left[S_m(t) - S_m\right]
 \left|\sbold_\initconf^{\phantom{\dag}} \right\rangle \right. \nonumber\\
& & \quad \quad +
\frac 1 2 
\left\langle \sbold_\initconf^{\phantom{\dag}}\right| \, 
4\left[S_m(t) - S_m\right]^2
\left|\sbold_\initconf^{\phantom{\dag}} \right\rangle\\
& & \quad \left. \quad -
\frac 1 2 
\left\langle \sbold_\initconf^{\phantom{\dag}}\right| \, 
2 \left[S_m(t) - S_m\right]
\left|\sbold_\initconf^{\phantom{\dag}} \right\rangle^2 
\right\}\qquad \\
&=& \exp\left\{ 
\left\langle \sbold_\initconf^{\phantom{\dag}}\right| \, 
2\left[ S_m(t) - S_m\right]^2
\left|\sbold_\initconf^{\phantom{\dag}} \right\rangle
\right\}\,.
\eea
\end{subequations}
The periodically oscillating terms giving the time dynamics of $\chi_{\sbold_\initconf,m}(t)$ average out in the long-time limit, so that we only need to evaluate $\left\langle \sbold_\initconf^{\phantom{\dag}}\right| \, 
2 S_m^2
\left|\sbold_\initconf^{\phantom{\dag}} \right\rangle$
 to arrive at Eq.~(4) given in the main text. 

In Fig.~\ref{fig:1} and~\ref{fig:2}, we show the ndRG results for the disorder-averaged many-body localization length for two values of $\alpha$. 
Although the present ndRG is a priori formulated for strong disorder, it agrees remarkably well to ED down to $W/J\lesssim 2$, and for $\alpha=0.5$ even over the entire range of disorder strengths. This good agreement gives confidence in the validity of the ndRG approach. 

\bibliographystyle{apsrev}
\bibliography{MBLIsing}



\end{document}